\documentclass[superscriptaddress,prl,twocolumn,nofootinbib]{revtex4}
\usepackage{amsmath,amssymb}
\usepackage{graphicx,epsfig}
\usepackage{slashed}
\usepackage{float,xcolor}

\def\beq{\begin{equation}}
\def\eeq#1{\label{#1}\end{equation}}
\def\eeqn{\end{equation}}
\def\beqa{\begin{eqnarray}}
\def\eeqa#1{\label{#1}\end{eqnarray}}
\def\eeqan{\end{eqnarray}}
\def\CR{\nonumber \\ }
\def\leqn#1{(\ref{#1})}
\newcommand{\centeron}[2]{{\setbox0=\hbox{#1}\setbox1=\hbox{#2}\ifdim
\wd1>\wd0\kern.5\wd1\kern-.5\wd0\fi \copy0
\kern-.5\wd0\kern-.5\wd1\copy1\ifdim\wd0>\wd1
                                   \kern.5\wd0\kern-.5\wd1\fi}}
\newcommand{\ltap}{\>\centeron{\raise.35ex\hbox{$<$}}
                           {\lower.65ex\hbox{$\sim$}}\>}
\newcommand{\gtap}{\>\centeron{\raise.35ex\hbox{$>$}}
                           {\lower.65ex\hbox{$\sim$}}\>}
\newcommand{\gsim}{\mathrel{\gtap}}
\newcommand{\lsim}{\mathrel{\ltap}}

\begin{document}

\title{Elastically Decoupling Dark Matter}

\author{Eric Kuflik}
\email{eak245@cornell.edu}
\author{Maxim Perelstein}
\email{mp325@cornell.edu}
\author{Nicolas Rey-Le Lorier}
\email{nr323@cornell.edu}
\author{Yu-Dai Tsai}
\email{yt444@cornell.edu}
\affiliation{Laboratory for Elementary Particle Physics, Cornell University,
Ithaca, NY 14850, USA}

\date{\today}

\begin{abstract}
We present a novel dark matter candidate, an Elastically Decoupling Relic (ELDER), which is a cold thermal relic whose present abundance is determined by the cross-section of its elastic scattering on Standard Model particles. The dark matter candidate is predicted to have a mass ranging from a few to a few hundred MeV, and an elastic scattering cross-section with electrons, photons and/or neutrinos in the $10^{−3}-1$~fb range.
\end{abstract}

%\pacs{14.60.Pq, 98.80.Cq, 98.70.Vc}

\maketitle
\section{Introduction}
It has now been firmly established that the Standard Model (SM) of particle physics must be extended to include new particle(s) to account for the observed dark matter (DM). %However, since our observational knowledge of dark matter is restricted to its gravitational interactions, microscopic properties of the DM particle are not well-constrained, and many microscopic models can fit the data. 
Many of the proposed dark matter candidates fall into a broad category of {\it thermal relics}, particles which were in thermal equilibrium with the hot SM particle plasma at some point in the early universe, and subsequently ``froze out" as the universe expanded and cooled~\cite{Lee:1977ua}. An attractive feature of this framework is its predictive power: the current abundance of the DM $\chi$
%~\cite{footnote1}  
 can be related to its microscopic properties, such as its mass and interaction cross sections. 

The type of interactions which determine the $\chi$ relic abundance can vary. The following three reactions will play a major role in the analysis of this paper:

\begin{itemize}
\item Elastic Scattering: $\chi$+SM $\leftrightarrow \chi$+SM, where ``SM" stands for any of the known Standard Model particles. 
\item Annihilation: $\chi+\chi \leftrightarrow$ SM+SM.
 \item Self-Annihilation: $\chi\chi\leftrightarrow  \overbrace{\chi
    \ldots \chi}^{n}$, with $n\geq 3$. (Specifically, we will focus on the case $n=3$.) 
\end{itemize}  
In the popular weakly-coupled massive particle (WIMP) paradigm, the relic abundance is entirely determined by the annihilation process. An alternative paradigm of self-interacting dark matter relies instead on self-annihilation~\cite{Carlson:1992fn}. %, with annihilation and elastic scattering processes decoupled at the time of freeze-out. 
Unfortunately, the dark matter predicted by this scheme is too light ($\lesssim 100$ eV) to be consistent with the observed large-scale structure~\cite{Carlson:1992fn,deLaix:1995vi}. Recently, an interesting variation has been proposed, dubbed the strongly interacting massive particle or SIMP~\cite{Hochberg:2014dra,Hochberg:2014kqa} (for extensions and variations, see \cite{Bernal:2015bla,Lee:2015gsa,Choi:2015bya,Bernal:2015ova,Bernal:2015xba}). In this model, the relic abundance is still set by self-annihilation, but the elastic scattering process is strong enough to sustain the thermal equilibrium between the SM and DM sectors until freeze-out occurs. In this case, the dark matter mass consistent with cosmological data is between an MeV and a GeV.%, and it is predicted to be of order 100~MeV (intriguingly close to the QCD confinement scale) when realized in a QCD-like theory~\cite{Hochberg:2014kqa}. 

In these and all other known examples, the DM relic abundance is set by processes that change the $\chi$ particle number. %This makes sense: such processes allow $\chi$ number density to evolve on its equilibrium trajectory as the universe cools, and freeze-out occurs when the annihilations decouple. 
In this Letter, we present a novel scenario in which the dark matter relic density is determined almost exclusively by the decoupling of the {\it elastic scattering}. We will refer to the dark matter candidate in this scenario as ``ELastically DEcoupling Relic", or ELDER. In a nutshell, the scenario works as follows. At high temperatures, when $\chi$ is relativistic, it is in thermal and chemical equilibrium with the SM plasma. As the universe cools to temperatures below the $\chi$ mass, the  $\chi$  equilibrium density drops exponentially, and the annihilation process quickly decouples. (This feature is the same as in the SIMP scenario.) The self-annihilation and elastic scattering processes are still active, and maintain thermal and chemical equilibrium (with zero chemical potential) between the two sectors. In the ELDER scenario, the elastic scattering decouples {\it first}, while the self-annihilation process is still active. (This is in contrast to the SIMP case~\cite{Hochberg:2014dra}, where the self-annihilation process is the first one to decouple.) After the decoupling of elastic scattering, the dark matter sector enters the so-called ``cannibalization" epoch~\cite{Carlson:1992fn}, in which the energy released by self-annihilation keeps it at an approximately constant temperature, even as the universe continues to expand. Eventually, the self-annihilation process also decouples, at which point the comoving number density of  $\chi$ is frozen. The near-constant temperature (and therefore density) of the DM in the cannibalization epoch means that the relic abundance of dark matter observed today is almost entirely fixed by the density of $\chi$'s at the beginning of this epoch, which in turn is fixed by the size of the elastic scattering cross section. 

We study the scenario outlined above using both simple estimates and detailed numerical solutions of the Boltzmann equations. We find that the observed dark matter abundance can be reproduced, and all theoretical and observational constraints can be satisfied, for $\chi$ masses between a few and a few hundred MeV, while the cross-section of elastic scattering between DM and SM particles (electrons, photons, and/or neutrinos) is of the order of $10^{-3}-1$ fb in the non-relativistic limit. DM candidates with such properties arise in simple and attractive theoretical extensions of the SM: for example, a hidden-sector DM can interact with the SM sector via a TeV-scale $Z^\prime$ with order-one gauge couplings to both sectors, or via a relatively light ($0.01-1$ GeV) dark photon with a kinetic mixing parameter $\epsilon\sim 10^{-8}$~\cite{long}.

\section{The elastically decoupling thermal relic}

\begin{figure}[t!]
\begin{center}
\includegraphics[width=8.5cm]{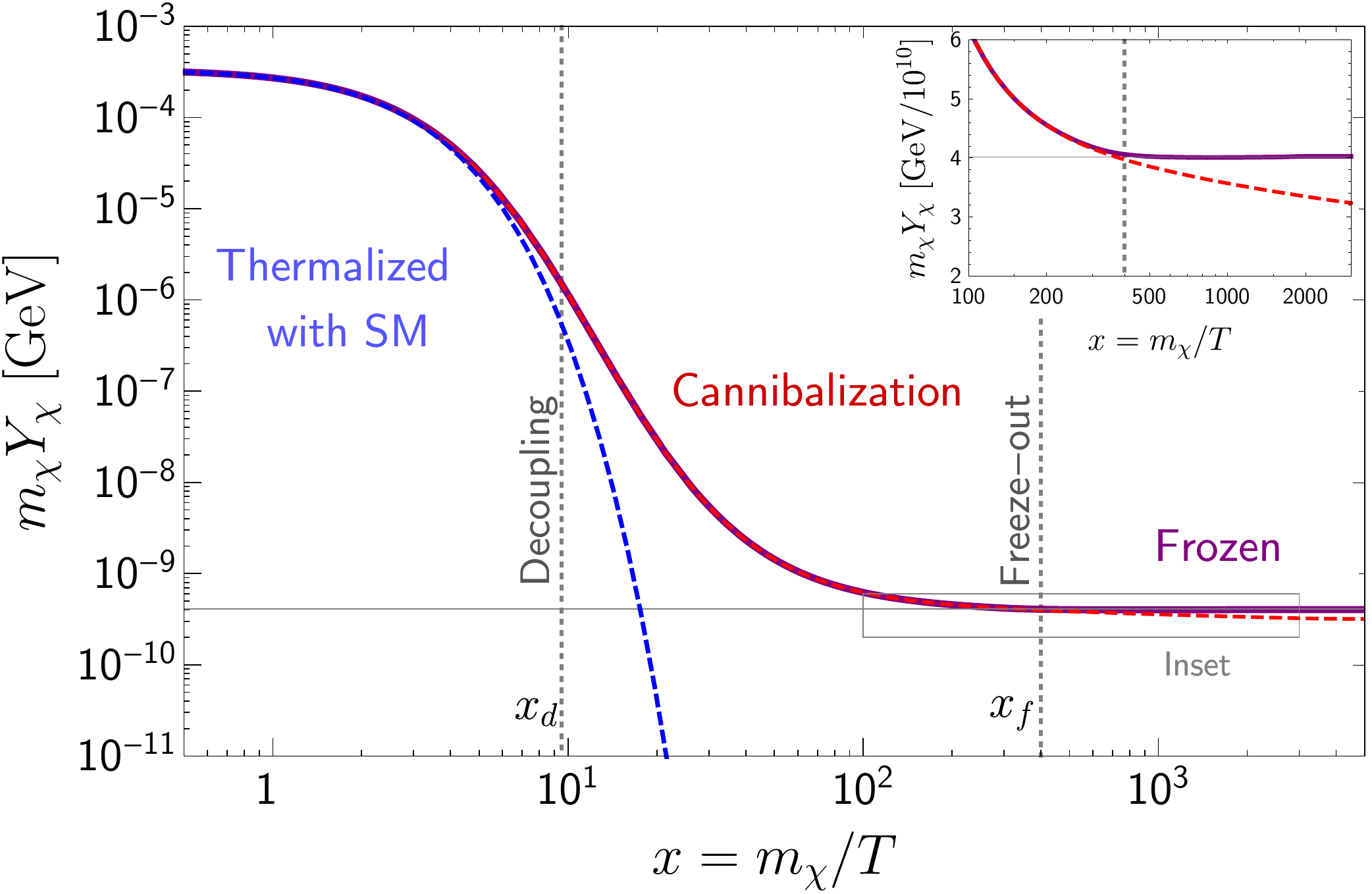}
\vspace{-2mm}
\caption{Dark matter yield, as a function of the SM plasma temperature $T$, for elastically decoupling dark matter with $m_\chi=10$ MeV, $\epsilon=8.5\times 10^{-9}$, and $\alpha=1$ (\textbf{{\color[rgb]{.5,0,.5}purple}}/solid line). For comparison, the dashed curves show the equilibrium yield assuming the DM and SM plasmas are in equilibrium (\textbf{{\color[rgb]{0,0,1}blue}}/dashed), and assuming the DM plasma is in chemical equilibrium with itself after decoupling (\textbf{{\color[rgb]{1,0,0}red}}/dashed).}
\label{fig:yield}
\end{center}
\end{figure}

The thermal history of the ELDER is summarized in Fig.~\ref{fig:yield}. At high temperatures, when $\chi$ is relativistic, it maintains thermal and chemical equilibrium with the SM plasma. As the universe cools, the temperature drops below the $\chi$ mass, and the subsequent thermal history is marked by two important events. First is ``decoupling", when the rate of elastic scattering becomes insufficient to maintain the DM and SM sectors in thermal contact. Second is ``freeze-out", at which point the rate of self-annihilation becomes insufficient to maintain chemical equilibrium in the DM sector, and the comoving dark matter density is frozen. Between these two events, chemical equilibrium within the DM sector are still maintained by self-annihilations, but the DM temperature $T^\prime$ is no longer equal to the SM temperature $T$. In this regime, the DM gas undergoes ``cannibalization": $3\to 2$ self-annihilations decrease the number density, but at the same time inject kinetic energy into the remaining gas. As the DM gas cannot exchange entropy with the SM sector at this time, its comoving entropy density is constant as the universe expands:
\beqa
a^3 s^\prime_\chi = a^3 \frac{m_\chi n_\chi}{T^\prime} ={\rm constant }\nonumber \\
 \Longrightarrow (T^{\prime})^{ 1/2} e^{-m_\chi/T^\prime} \propto T^3
\eeqa{cannibalentropy}
where $a \propto T^{-1}$ is the FRW scale-factor. As a result,  $T^\prime$ decreases much slower than $T$ as the universe expands: 
\beq
T^\prime \approx \frac{T_d}{1+3x_d^{-1} \log T_d/T}\,,
\eeq{cannibal}
where $x_d\equiv m_\chi/T_d$ and $T_d$ is temperature at which (elastic) decoupling occurs. The comoving DM number density, plotted in  Fig.~\ref{fig:yield}, changes very slowly during the cannibalization regime.

Let $T_f^\prime$ denote the DM temperature at freeze-out. Since the comoving entropies of the DM and SM sectors are separately conserved in the cannibalization epoch, the DM number density at freeze-out is given by
\beq
n_f^\prime = \frac{\rho_f^\prime}{m_\chi} = \frac{s_f^\prime T_f^\prime}{m_\chi} = \frac{s_d^\prime}{x_f^\prime}\frac{s_f}{s_d},
\eeq{nfprime}
where $x_f^\prime=m_\chi/T_f^\prime$, $s_d$ and $s_d^\prime$ are the entropy densities of the SM and DM sectors at decoupling, and  $s_f$ and $s_f^\prime$ 
are the same quantities at freeze-out. The DM number density today is
\beq
n_0 = \frac{s_0}{s_f}n_f^\prime = \frac{s_d^\prime}{s_d} \frac{s_0}{x_f^\prime},
\eeq{n0}
where $s_0$ %$\approx 7.04 n_\gamma^{\rm CMB}$ 
is the current entropy density. Since the dark matter is non-relativistic at $T_d$,
\beq
\Omega_\chi = \frac{45}{2^{5/2}\pi^{3/2}}\,\left( \frac{m_\chi s_0}{\rho_c} \right)\,\left( \frac{g_\chi}{g_{*d}}\right) \, \frac{x_d^{5/2}e^{-x_d}}{x_f^\prime},
\eeq{omega1}
where $\rho_c$ is the critical density ($s_0/\rho_c\approx $ 0.60 eV$^{-1}$), $g_\chi$ is the number of degrees of freedom in the $\chi$ field ({\it e.g.} 2 for complex scalar and 4 for Dirac fermion), and $g_{*d}$ is the effective number of relativistic SM degrees of freedom at decoupling. Hence, the relic abundance is exponentially sensitive to the temperature at which the elastic scattering processes decouple. 

In order to determine the temperatures at decoupling, $x_d$,  and at freeze-out, $x_f^\prime$, we parametrize the elastic scattering and self-annihilation cross-sections in the non-relativistic limit as
\beq
\lim_{T\rightarrow 0} \langle \sigma_{\rm el}v\rangle \equiv \frac{\epsilon^2}{m_\chi^2},~~~\lim_{T\rightarrow 0} \langle \sigma_{3\to2}v^2\rangle \equiv \frac{\alpha^3}{m_\chi^5},
\eeq{alpha_eps_defs} 
where $\sigma_{\rm el}$ is the cross-section of elastic scattering, averaged over SM species that are relativistic at $T\sim m_\chi$. At $T<m_\chi$, the equilibrium density of DM particles drops exponentially as $n^{\rm eq}_\chi\sim (m_\chi T)^{3/2} e^{-m_\chi/T}$. The self-annihilation process which maintains chemical equilibrium in the DM gas releases kinetic energy, at a per-particle rate of %$\dot{K}_\chi = m_\chi \left.{ (\dot{n}}/{n})\right|_{\mu_\chi =0} =  m_\chi^2 H T^{-1}.$
\beq
\dot{K}_\chi = m \left.\frac{ \dot{n}}{n}\right|_{\mu_\chi =0} \simeq  -m_\chi^2 H T^{-1}.
\eeq{hubbleloss} 
Elastic scattering processes transfer this excess kinetic energy to the SM gas at a rate 
\beq
\dot{K}_\chi \sim \Gamma_{\rm el} v^2_\chi T \sim T^5 \epsilon^2/m_\chi^3,
\eeq{elasticloss}
where $\Gamma_{\rm el} =  n_{\rm SM} \langle\sigma_{\rm el} v\rangle$ is the rate at which each $\chi$ scatters elastically off the SM gas. The decoupling occurs when the DM-to-SM energy transfer can no longer keep up with the kinetic energy production; equating Eq.~\leqn{hubbleloss} with Eq.~\leqn{elasticloss},
\beq
x_d \sim \epsilon^{1/2}m_\chi^{-1/4} M_{\rm Pl}^{1/4}.
\eeq{Td}
%The relic abundance,  given in (\ref{omega1}), is exponentially sensitive to this value. 

Freeze-out occurs when the rate of self-annihilations is no longer sufficient to maintain chemical equilibrium, $(n_\chi^{\rm eq})^2\langle \sigma_{3\to 2} v_\chi\rangle \sim \dot{n}_\chi^{\rm eq}/n_\chi^{\rm eq}$, which yields
\beq
x^\prime_f \sim \frac{3}{4} \log\left( \frac{M_{\rm Pl}}{m_\chi} \right) -\frac{x_d}{2} + \frac{9}{4} \log \alpha.
\eeq{Tf}
For DM mass in the MeV$-$GeV range, the relic density can be conveniently approximated as
\beq
\Omega_\chi \sim \frac{10^6 m_{\rm MeV} \exp(-10\epsilon_{-9}^{1/2} m_{\rm MeV}^{-1/4})}{1+0.07\log\alpha},
\eeq{Omega_master}
where $\epsilon_{-9} \equiv \epsilon/10^{-9}$ and $m_{\rm MeV}\equiv m_\chi/(1~{\rm MeV})$.
As emphasized in the Introduction, the relic density is controlled by the strength of the elastic scattering, $\epsilon$, with only weak, logarithmic, dependence on the strength of the number-changing self-annihilation process $\alpha$. This is the unique feature of the ELDER scenario. 

The ELDER mechanism is only possible if the self-annihilation process maintains the DM gas in chemical equilibrium until {\it at least} the temperature $T_d$, requiring
\beqa
\alpha \gsim \alpha_{\rm min} &\simeq& \frac{10^{-5} x_d^{7/3} m_{\rm MeV}}{\Omega_\chi^{2/3}} \CR
&\approx& 0.015 m_{\rm MeV} \left( 1+0.16 \log m_{\rm MeV} \right). 
\eeqa{alpha_bound}
Numerical solutions to the Boltzmann equations (see below) indicate that Eq.~\leqn{alpha_bound} somewhat underestimates the lower bound of the ``pure ELDER" region: for ${\alpha\lesssim {\rm~a~few~} \times\alpha_{\rm min}}$, both self-annihilation and elastic scattering are important. For even lower $\alpha$,   
 freeze-out occurs before the elastic scattering decouples; this is precisely the SIMP scenario of~\cite{Hochberg:2014dra,Hochberg:2014kqa}. Together with perturbativity and unitarity constraints on the self-annihilation cross section, which can be estimated as $\alpha\lesssim 4\pi$, this bound imposes an upper bound on the DM mass. For a ``pure ELDER,'' this implies $m_\chi \lesssim 100$ MeV. A lower bound of $m_\chi \gtrsim$ a few MeV is imposed by observational constraints, see below. In the allowed mass range, the correct relic density is obtained for $\epsilon$ ranging between $10^{-9}$ and $10^{-7}$, while $\alpha\sim10^{-2}-10$.      

A potential concern in the elastic decoupling scenario is its naturalness: if small changes in $\epsilon$ lead to huge changes in the relic density, it would be difficult to conceive of a reason why $\Omega_\chi\sim 1$ in the observed universe. To quantify this issue, we estimate 
\beq
\frac{\partial \log \Omega_\chi}{\partial \log \epsilon} \approx 7 + \frac{1}{2}\log m_{\rm MeV}. 
\eeq{fine_tuning}   
An order-of-magnitude change in the relic density requires a $20-30$\% change in $\epsilon$. We conclude that only a mild amount of tuning is required to obtain $\Omega_\chi \sim 1$. %\EK{I think this paragraph can be removed is space is needed.}

\section{The Boltzmann Equations}

\begin{figure}[t!]
\begin{center}
\includegraphics[width=8.7cm]{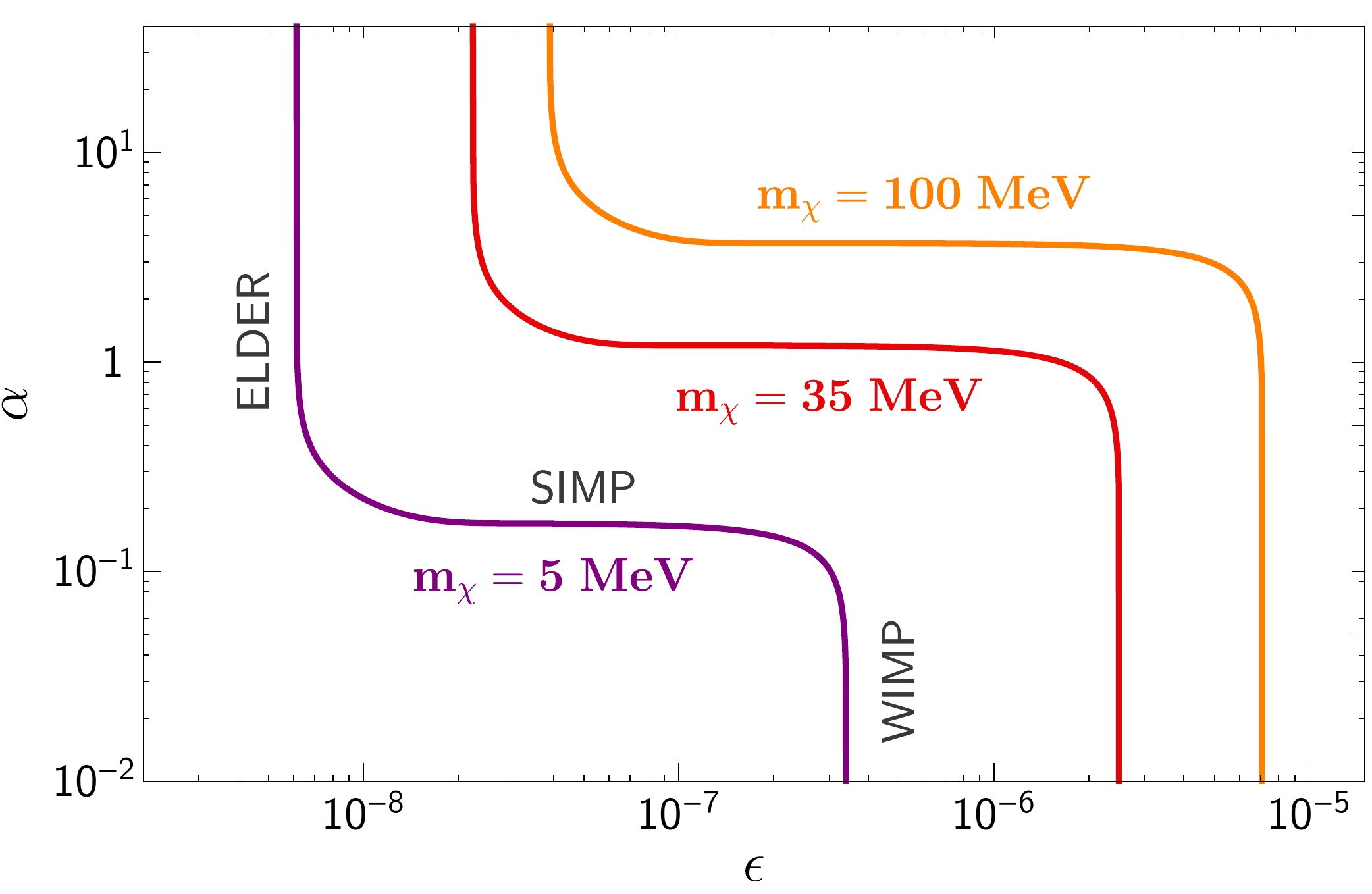}
\vspace{-2mm}
\caption{Regions of parameters corresponding to the observed relic density. For each mass, the vertical section of the line of the left/top corresponds to the elastically decoupling relic (ELDER) scenario proposed in this paper; the horizontal line to the SIMP scenario; and the vertical section on the right/bottom to the WIMP scenario.}
\label{fig:epsvsalpha}
\end{center}
\end{figure}

The starting point of the analysis is the microscopic Boltzmann equation for the phase-space density of the DM particle $\chi$, with collision terms describing elastic scattering, $\chi\gamma\to\chi\gamma$; annihilation $\chi\chi\leftrightarrow\gamma\gamma$; and self-annihilation $\chi\chi\leftrightarrow\chi\chi\chi$. (For concreteness, we assume that the dominant DM coupling to the SM is via photons; couplings to $e^\pm$ or $\nu$ would produce similar results.)  Since the $\chi$ velocities follow thermal distribution at all times, %At high temperatures, This is enforced by the elastic scattering $\chi\chi\to\chi\chi$, which remain active throughout freeze-out. %; after this process decouples, the velocities redshift uniformly and the shape of the distribution is unaffected. 
the microscopic Boltzmann equation reduces to two integro-differential equations for the DM number density $n_\chi(t)$, 
\beq
 \frac{d{n}_{\chi}}{dt} +3H n_{\chi}= -\langle \sigma_{3\to 2}  v^2 \rangle(n^3_{\chi} - n_{\chi}^2 n_{\chi}^{\rm eq}) + \ldots,
\eeq{boltz_num}
and energy density $\rho_\chi(t)$,
\beq 
 \frac{d{\rho}_{\chi}}{dt} +3H ( \rho_{\chi} + P_{\chi}  ) = \langle  \sigma_{\rm el} v \cdot \delta E \rangle n_{\chi} n_{\rm \gamma}^{\rm eq}+\ldots,
 \eeq{boltz_rho}
 where 
% \beq\def\arraystretch{1.5}\begin{array}{r}
% \langle  \sigma v \cdot \delta E \rangle_{\rm kin} = \frac{1}{n_{\chi}^{\rm eq} n_{\rm \gamma}^{\rm eq} } \int d\Pi_{\chi_1} d\Pi_{\gamma_1} d\Pi_{\chi_2} d\Pi_{\gamma_2} (2\pi )^4 \delta ^4\left(p\right)   \\
%\times (E_{\chi_2} - E_{\chi_1}) e^{-E_{\chi_1}/T'- E_{\gamma_1}/T} \left|M\right|^2 .
%  \end{array}\eeq{energy_scattering}
 \beqa
 \langle  \sigma_{\rm el} v \cdot \delta E \rangle &=& \frac{1}{n_{\chi}^{\rm eq} n_{\rm \gamma}^{\rm eq} } \int d\Pi_{\chi_1} d\Pi_{\gamma_1} d\Pi_{\chi_2} d\Pi_{\gamma_2} (2\pi )^4 \delta ^4\left(p\right) \nonumber  \\
&& \times (E_{\chi_2} - E_{\chi_1}) e^{-E_{\chi_1}/T'- E_{\gamma_1}/T} \left|\mathcal{M}\right|^2 ,
\eeqa{energy_scattering}
 and $d\Pi_{i} = g_i d^3p_1/(2\pi)^{3} $. Here, the dots denote the annihilation terms; these are unimportant in the ELDER regime, but are nevertheless fully included in the numerical analysis, as will be described in detail in Ref.~\cite{long}. 

The numerical solution for the evolution of the DM yield, $Y_\chi \equiv n_\chi/s$, in the ELDER scenario is shown in Fig.~\ref{fig:yield}. The three stages of the DM evolution (thermal equilibrium with the SM, cannibalization, and freeze-out) are clearly visible. The yield evolves very slowly in the cannibalization stage, due to slow evolution of the DM temperature (for the parameters in Fig.~\ref{fig:yield}, $T^\prime_f\approx 0.3 T_d$, while $T_f\approx 0.025 T_d$). As a result, the final DM abundance is approximately independent of when freeze-out occurs, and hence of the self-annihilation cross-section. 

This feature is further illustrated in Fig.~\ref{fig:epsvsalpha}, which shows the regions of parameter space where the observed DM density is reproduced. For fixed $m_\chi$, the ELDER scenario corresponds to the narrow vertical region of approximately constant $\epsilon$, while $\alpha$ can take any value above a certain lower cutoff; these features are consistent with the estimates in Eqs.~\leqn{Omega_master} and \leqn{alpha_bound}. For smaller values of $\alpha$, self-annihilations freeze out before elastic scattering decouples, and the relic density is fixed by the strength of the self-annihilation process, $\alpha$, and is independent of $\epsilon$ as long as it is large enough. The resulting horizontal region corresponds precisely to the SIMP scenario proposed in~\cite{Hochberg:2014dra}. Finally, if $\epsilon$ becomes too large, annihilations become important, and since $\epsilon$ controls the annihilation cross-section, another vertical region occurs. This corresponds to the canonical WIMP scenario (or the ``WIMPless"  regime~\cite{Feng:2008ya}). The numerical study clearly establishes the presence of the novel elastic decoupling scenario. In addition, it establishes precise boundaries of the different regimes, and traces out in detail the transition regions where two types of interactions play an equally important role in setting the relic density. 

\section{Constraints} 

\begin{figure}[t!]
\begin{center}
\includegraphics[width=8.8cm]{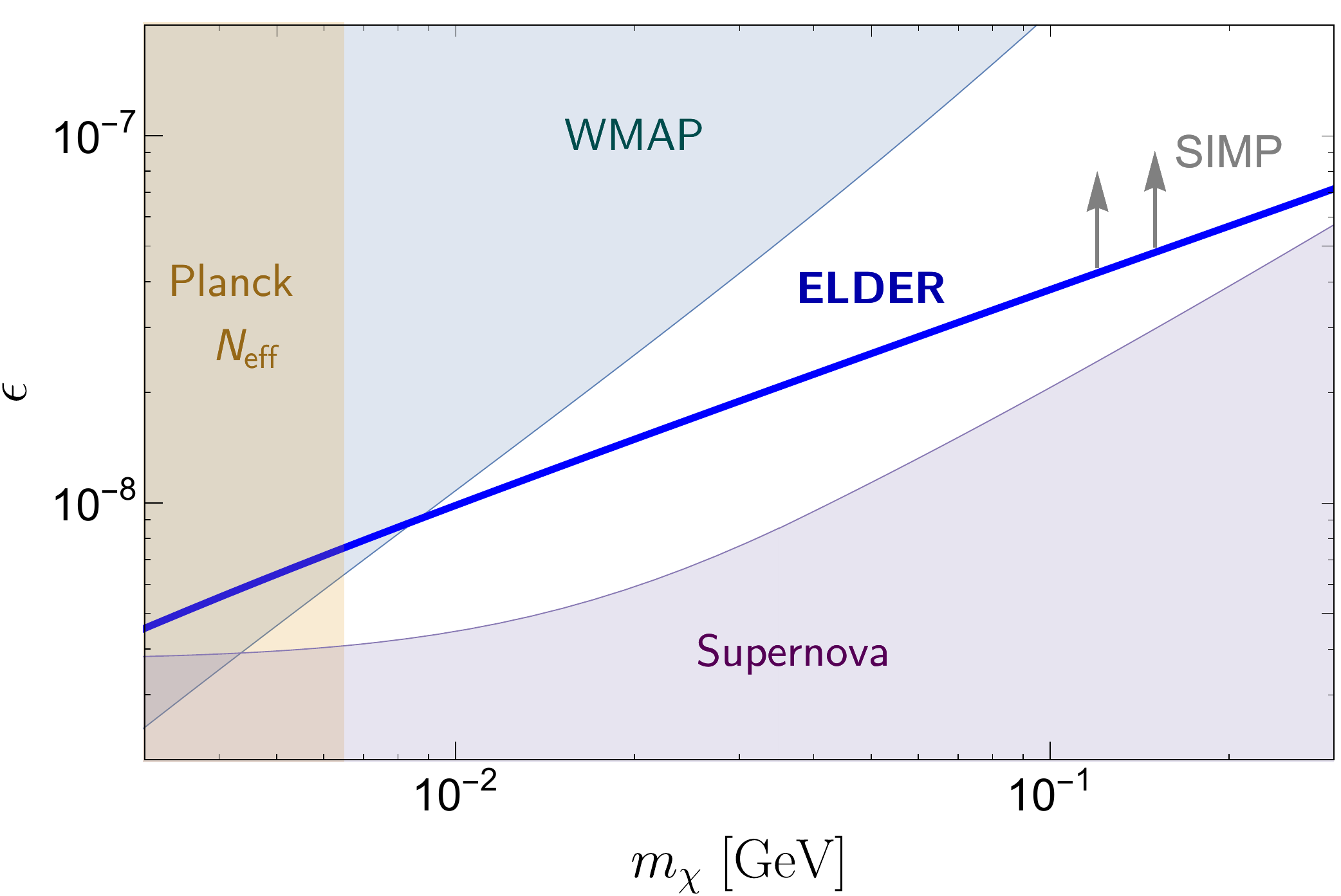}
\vspace{-2mm}
\caption{Constraints on $\epsilon$ vs. $m_{\chi}$,  from dark matter couplings to photons. The blue line corresponds to the ELDER scenario while the region above it corresponds to the SIMP scenario.  Also shown are the exclusion limits from: supernova cooling (purple region); CMB constraints on DM annihilations into photons before recombination (blue region); and modification to $N^\nu_{\rm eff}$ from DM decoupling (red region).
}
\label{fig:epsvsmass}
\end{center}
\end{figure}

Since ELDER dark matter has mass and coupling strengths similar to the case of SIMPs, the same set of observational constraints is relevant for both scenarios. The strongest constraints on the strength of the $\chi$ interactions with the SM are summarized in Fig.~\ref{fig:epsvsmass}. 

%Consistently with the discussion above, the constraints shown here assume that $\chi$ couples only to photons, and that elastic scattering and annihilation cross sections approach non-vanishing constant values in the non-relativistic limit. In most cases, modifying these assumptions would further weaken the bounds. 

% on the strength of DM-SM interactions relevant for our scenario are summarized in Fig.~\ref{fig:epsvsmass}. All constraints are somewhat model-dependent, and we use the most conservative interpretation in all cases. In a specific model, the bounds may be different than presented here. Here we assume that the dark-matter only couples to photons. Elastic decoupling from electrons or neutrinos will generally weaken the bounds discussed below \EK{Are we being conservative or aggressive?}
 
In the relevant range of $\epsilon$, the reaction $\gamma\gamma\to\chi\chi$ in the core of the supernova SN1987A would lead to energy loss rate inconsistent with observations, unless the produced $\chi$ particles become trapped in the core~\cite{Ellis:1988aa,Grifols:1988fw,Raffelt:1996wa,Dreiner:2003wh}. Since trapping is due to elastic scattering of $\chi$ on photons in the supernova core, this constraint places a {\it lower} bound on $\epsilon$. The value predicted by the ELDER scenario satisfies this bound throughout the relevant mass range. 
The bound can be further weakened if $\chi$ couples to $e^-$ or $\nu$ instead of $\gamma$, as their higher density in the supernova core implies a smaller mean free path for the same value of $\epsilon$. 

Cosmic Microwave Background (CMB) measurements limit the rate of DM annihilation into SM particles before recombination, which can distort the energy spectrum of the CMB~\cite{Finkbeiner:2011dx,Madhavacheril:2013cna,Slatyer:2015jla}. In our case, the relevant process is $\chi\chi\to\gamma\gamma$. The cross-section of this process in the non-relativistic regime is obtained from Eq.~\leqn{alpha_eps_defs}, which implies that annihilation occurs in $s$-wave. The WMAP results~\cite{Hinshaw:2012aka} place an upper bound on $\epsilon$ shown in Fig.~\ref{fig:epsvsmass}. Again, the coupling predicted by the ELDER scenario is consistent with this bound. Null results of searches for anomalous high-energy photons from dark matter annihilation in the Milky Way or its dwarf satellites can also be used to place an upper bound on $\epsilon$~\cite{Essig:2013goa}. The bound is similar to the one implied by the WMAP data, and we do not show it in Fig.~\ref{fig:epsvsmass}. Note that if $\chi\chi\to\gamma\gamma$ annihilation occurred in $p$-wave instead, the cross-sections relevant for both CMB and indirect searches would be severely suppressed relative to that at the time of dark matter decoupling, due to lower $\chi$ velocities, and the bounds would be even weaker. These bounds would also be completely eliminated if $\chi$ couples only to neutrinos.

If the ELDER decoupling occurs after the neutrinos are decoupled from the SM plasma, it can affect the temperature ratio $T_\nu/T_\gamma$, resulting in a non-standard value of $N^\nu_{\rm eff}$ measured in CMB observations. This places a lower bound on the DM mass of a few MeV, with the exact number depending on $g_\chi$: for example, $m_\chi\gsim 6.5$ MeV for a complex scalar $\chi$ coupled only to $\gamma$~\cite{Boehm:2013jpa}. This bound can be avoided if $\chi$ is coupled to both $\nu$ and $e/\gamma$, since in this case reheating due to ELDER decoupling does not change the ratio $T_\nu/T_\gamma$. (The region $m_\chi\lsim 1$ MeV is also constrained by the Big Bang Nucleosynthesis bound on the number of relativistic degrees of freedom.) %This bound could be avoided if $\chi$ is a real scalar; however, there is no known symmetry mechanism for ensuring stability of such a state. 
In summary, the ELDER scenario is consistent with all constraints provided that $m_\chi \gsim$ a few MeV, even with the most stringent interpretation of the observational bounds.

If the DM couples to electrons, additional signatures arise in direct detection experiments searching for electron recoils~\cite{Essig:2011nj}, as well as collider searches for $e^+e^-\to\chi\chi\gamma$~\cite{Birkedal:2004xn,Fox:2011fx,Chae:2012bq}. Current direct-detection bounds from XENON10~\cite{Essig:2012yx} are not yet sensitive to $\epsilon$ in the range predicted by the ELDER scenario, while the collider bounds from LEP-2 depend strongly on the mass of the particle mediating the DM-SM scattering, and cannot be used to put robust constrains on $\epsilon$. Interestingly, proposed dedicated germanium or silicon-based electron-recoil direct detection experiments~\cite{Essig:2015cda} and superconducting detectors~\cite{Hochberg:2015pha} may have the sensitivity to directly probe the ELDER scenario.  

Finally, the strong $3\to 2$ self-annihilations required in the ELDER scenario generically imply a large contribution to  $\chi \chi \rightarrow \chi \chi $ elastic self-scattering.  The elastic self-scattering cross-section at low velocities, $v_\chi\sim 10^{-3}$, is constrained by observations of the Bullet Cluster~\cite{Clowe:2003tk,Markevitch:2003at,Randall:2007ph} and halo shapes~\cite{Rocha:2012jg,Zavala:2012us,Peter:2012jh}:
\beq
\frac{\sigma_{\chi\chi\to\chi\chi}}{m_\chi} \lsim 1~{\rm cm}^2/{\rm g}.
\eeq{bullet} 
Note that a self-scattering cross section in the $0.1-1$ cm$^2$/g range~\cite{ Zavala:2012us,Vogelsberger:2012ku,Rocha:2012jg,Peter:2012jh}, consistent with this bound, could reconcile the N-body simulation results with the observed small-scale structure, providing an additional motivation for self-interacting DM candidates (see for instance,~\cite{Spergel:1999mh,deBlok:2009sp,BoylanKolchin:2011de,Kaplinghat:2015aga}). The precise relation between the elastic self-scattering and self-annihilation cross-sections is model-dependent. Generically, one might expect that
\beq
\sigma_{\chi\chi\to\chi\chi}=a^2 \frac{\alpha^2}{m_\chi^2},
\eeq{22xsec}
where $a$ is an order-one constant. Consistency of the ELDER scenario with the bound of Eq.~\leqn{bullet} requires $a \lsim 0.01-0.1$ (depending on $m_\chi$). However, if the self-annihilation and self-scattering cross-sections both vanish at threshold, and are therefore velocity suppressed, these bounds may be alleviated~\cite{long}.

~\\
{\em Acknowledgments ---} We are grateful to Yonit Hochberg, Hitoshi Murayama, Aaron Pierce, Josh Ruderman, and Yue Zhao for 
  useful discussions. This work is
  supported by the U.S. National Science Foundation through grant
  PHY-1316222. EK is supported by a Hans Bethe Postdoctoral Fellowship at Cornell. YT is also supported by Taiwan Study Abroad Scholarship.

~\\


\begin{thebibliography}{99}


%\cite{Lee:1977ua}
\bibitem{Lee:1977ua}
B.~W.~Lee and S.~Weinberg,
%``Cosmological Lower Bound on Heavy Neutrino Masses,''
Phys.\ Rev.\ Lett.\ {\bf 39} (1977) 165.
%doi:10.1103/PhysRevLett.39.165
%%CITATION = doi:10.1103/PhysRevLett.39.165;%%<br /> 861 citations counted in INSPIRE as of 08 Dec 2015



%\bibitem{footnote1}
%For simplicity, we will restrict our attention to the case where all of dark matter is made out of the same particle species, $\chi$.

\bibitem{Carlson:1992fn} 
  E.~D.~Carlson, M.~E.~Machacek and L.~J.~Hall,
  %``Self-interacting dark matter,''
  Astrophys.\ J.\  {\bf 398}, 43 (1992).
%  doi:10.1086/171833
  %%CITATION = doi:10.1086/171833;%%

\bibitem{deLaix:1995vi} 
  A.~A.~de Laix, R.~J.~Scherrer and R.~K.~Schaefer,
  %``Constraints of selfinteracting dark matter,''
  Astrophys.\ J.\  {\bf 452}, 495 (1995)
%  doi:10.1086/176322
  [astro-ph/9502087].
  %%CITATION = doi:10.1086/176322;%%

\bibitem{Hochberg:2014dra} 
  Y.~Hochberg, E.~Kuflik, T.~Volansky and J.~G.~Wacker,
  %``Mechanism for Thermal Relic Dark Matter of Strongly Interacting Massive Particles,''
  Phys.\ Rev.\ Lett.\  {\bf 113}, 171301 (2014)
  [arXiv:1402.5143 [hep-ph]].
  %%CITATION = ARXIV:1402.5143;%%

\bibitem{Hochberg:2014kqa} 
  Y.~Hochberg, E.~Kuflik, H.~Murayama, T.~Volansky and J.~G.~Wacker,
  %``Model for Thermal Relic Dark Matter of Strongly Interacting Massive Particles,''
  Phys.\ Rev.\ Lett.\  {\bf 115}, no. 2, 021301 (2015)
  [arXiv:1411.3727 [hep-ph]].
  %%CITATION = ARXIV:1411.3727;%%
  
%\cite{Bernal:2015bla}
\bibitem{Bernal:2015bla} 
  N.~Bernal, C.~Garcia-Cely and R.~Rosenfeld,
  %``WIMP and SIMP Dark Matter from the Spontaneous Breaking of a Global Group,''
  JCAP {\bf 1504}, no. 04, 012 (2015)
%  doi:10.1088/1475-7516/2015/04/012
  [arXiv:1501.01973 [hep-ph]].
  %%CITATION = doi:10.1088/1475-7516/2015/04/012;%%
  %8 citations counted in INSPIRE as of 11 Dec 2015
  
  %\cite{Lee:2015gsa}
\bibitem{Lee:2015gsa} 
  H.~M.~Lee and M.~S.~Seo,
  %``Communication with SIMP dark mesons via Z′ -portal,''
  Phys.\ Lett.\ B {\bf 748}, 316 (2015)
%  doi:10.1016/j.physletb.2015.07.013
  [arXiv:1504.00745 [hep-ph]].
  %%CITATION = doi:10.1016/j.physletb.2015.07.013;%%
  %5 citations counted in INSPIRE as of 11 Dec 2015

%\cite{Choi:2015bya}
\bibitem{Choi:2015bya} 
  S.~M.~Choi and H.~M.~Lee,
  %``SIMP dark matter with gauged Z$_{3}$ symmetry,''
  JHEP {\bf 1509}, 063 (2015)
%  doi:10.1007/JHEP09(2015)063
  [arXiv:1505.00960 [hep-ph]].
  %%CITATION = doi:10.1007/JHEP09(2015)063;%%
  %5 citations counted in INSPIRE as of 11 Dec 2015
  
 %\cite{Bernal:2015ova}
\bibitem{Bernal:2015ova} 
  N.~Bernal, X.~Chu, C.~Garcia-Cely, T.~Hambye and B.~Zaldivar,
  %``Production Regimes for Self-Interacting Dark Matter,''
  arXiv:1510.08063 [hep-ph].
  %%CITATION = ARXIV:1510.08063;%%
  %1 citations counted in INSPIRE as of 11 Dec 2015 
  
%\cite{Bernal:2015xba}
\bibitem{Bernal:2015xba} 
  N.~Bernal and X.~Chu,
  %``$Z_2$ SIMP Dark Matter,''
  arXiv:1510.08527 [hep-ph].
  %%CITATION = ARXIV:1510.08527;%%

  
\bibitem{long}
E.~Kuflik, M.~Perelstein, N.~Rey-Le Lorier, and Y.-D.~Tsai, 
%{\it ``WIMPs, SIMPs and ELDERs: The Phase Diagram of Cold Thermal Relic Dark Matter"}, 
in preparation.

\bibitem{Feng:2008ya} 
  J.~L.~Feng and J.~Kumar,
  %``The WIMPless Miracle: Dark-Matter Particles without Weak-Scale Masses or Weak Interactions,''
  Phys.\ Rev.\ Lett.\  {\bf 101}, 231301 (2008)
%  doi:10.1103/PhysRevLett.101.231301
  [arXiv:0803.4196 [hep-ph]].
  %%CITATION = doi:10.1103/PhysRevLett.101.231301;%%
  
  %\cite{Ellis:1988aa,Dreiner:2003wh,Grifols:1988fw}
\bibitem{Ellis:1988aa} 
  J.~R.~Ellis, K.~A.~Olive, S.~Sarkar and D.~W.~Sciama,
  %``Low Mass Photinos and Supernova {SN1987A},''
  Phys.\ Lett.\ B {\bf 215}, 404 (1988).
%  doi:10.1016/0370-2693(88)91456-6
  %%CITATION = doi:10.1016/0370-2693(88)91456-6;%%
  %34 citations counted in INSPIRE as of 11 Dec 2015
  %\cite{Grifols:1988fw}
\bibitem{Grifols:1988fw} 
  J.~A.~Grifols, E.~Masso and S.~Peris,
  %``Photinos From Gravitational Collapse,''
  Phys.\ Lett.\ B {\bf 220}, 591 (1989).
%  doi:10.1016/0370-2693(89)90792-2
  %%CITATION = doi:10.1016/0370-2693(89)90792-2;%%
  %24 citations counted in INSPIRE as of 11 Dec 2015
  %\cite{Raffelt:1996wa}
\bibitem{Raffelt:1996wa} 
  G.~G.~Raffelt,
 {\it ``Stars as laboratories for fundamental physics : The astrophysics of neutrinos, axions, and other weakly interacting particles,''}
  Chicago, USA: Univ. Pr. (1996) 664 p
  %23 citations counted in INSPIRE as of 11 Dec 2015
  
 %\cite{Dreiner:2003wh}
\bibitem{Dreiner:2003wh} 
  H.~K.~Dreiner, C.~Hanhart, U.~Langenfeld and D.~R.~Phillips,
  %``Supernovae and light neutralinos: SN1987A bounds on supersymmetry revisited,''
  Phys.\ Rev.\ D {\bf 68}, 055004 (2003)
%  doi:10.1103/PhysRevD.68.055004
  [hep-ph/0304289].
  %%CITATION = doi:10.1103/PhysRevD.68.055004;%%
  %46 citations counted in INSPIRE as of 11 Dec 2015 
  
  
%\cite{Finkbeiner:2011dx}
\bibitem{Finkbeiner:2011dx} 
  D.~P.~Finkbeiner, S.~Galli, T.~Lin and T.~R.~Slatyer,
  %``Searching for Dark Matter in the CMB: A Compact Parameterization of Energy Injection from New Physics,''
  Phys.\ Rev.\ D {\bf 85}, 043522 (2012)
%  doi:10.1103/PhysRevD.85.043522
  [arXiv:1109.6322 [astro-ph.CO]].
  %%CITATION = doi:10.1103/PhysRevD.85.043522;%%
  %71 citations counted in INSPIRE as of 10 Dec 2015
%\cite{Madhavacheril:2013cna,Slatyer:2015jla}
\bibitem{Madhavacheril:2013cna}
M.~S.~Madhavacheril, N.~Sehgal and T.~R.~Slatyer,
%``Current Dark Matter Annihilation Constraints from Cmb and Low-Redshift Data,''
Phys.\ Rev.\ D {\bf 89} (2014) 103508
%doi:10.1103/PhysRevD.89.103508
[arXiv:1310.3815 [astro-ph.CO]].
%%CITATION = doi:10.1103/PhysRevD.89.103508;%%<br /> 64 citations counted in INSPIRE as of 10 Dec 2015

 
  
%\cite{Slatyer:2015jla}
\bibitem{Slatyer:2015jla}
T.~R.~Slatyer,
%``Indirect Dark Matter Signatures in the Cosmic Dark Ages I. Generalizing the Bound on S-Wave Dark Matter Annihilation from Planck,''
arXiv:1506.03811 [hep-ph].
%%CITATION = ARXIV:1506.03811;%%<br /> 11 citations counted in INSPIRE as of 10 Dec 2015

%\cite{Hinshaw:2012aka}
\bibitem{Hinshaw:2012aka} 
  G.~Hinshaw {\it et al.} [WMAP Collaboration],
  %``Nine-Year Wilkinson Microwave Anisotropy Probe (WMAP) Observations: Cosmological Parameter Results,''
  Astrophys.\ J.\ Suppl.\  {\bf 208}, 19 (2013)
%  doi:10.1088/0067-0049/208/2/19
  [arXiv:1212.5226 [astro-ph.CO]].
  %%CITATION = doi:10.1088/0067-0049/208/2/19;%%
  %1809 citations counted in INSPIRE as of 11 Dec 2015
  
  %\cite{Essig:2013goa}
\bibitem{Essig:2013goa} 
  R.~Essig, E.~Kuflik, S.~D.~McDermott, T.~Volansky and K.~M.~Zurek,
  %``Constraining Light Dark Matter with Diffuse X-Ray and Gamma-Ray Observations,''
  JHEP {\bf 1311}, 193 (2013)
%  doi:10.1007/JHEP11(2013)193
  [arXiv:1309.4091 [hep-ph]].
  %%CITATION = doi:10.1007/JHEP11(2013)193;%%
  %43 citations counted in INSPIRE as of 10 Dec 2015

\bibitem{Boehm:2013jpa} 
  C.~Boehm, M.~J.~Dolan and C.~McCabe,
  %``A Lower Bound on the Mass of Cold Thermal Dark Matter from Planck,''
  JCAP {\bf 1308}, 041 (2013)
%  doi:10.1088/1475-7516/2013/08/041
  [arXiv:1303.6270 [hep-ph]].
  %%CITATION = doi:10.1088/1475-7516/2013/08/041;%%

\bibitem{Essig:2011nj} 
  R.~Essig, J.~Mardon and T.~Volansky,
  %``Direct Detection of Sub-GeV Dark Matter,''
  Phys.\ Rev.\ D {\bf 85}, 076007 (2012)
%  doi:10.1103/PhysRevD.85.076007
  [arXiv:1108.5383 [hep-ph]].
  %%CITATION = doi:10.1103/PhysRevD.85.076007;%%

\bibitem{Birkedal:2004xn} 
  A.~Birkedal, K.~Matchev and M.~Perelstein,
  %``Dark matter at colliders: A Model independent approach,''
  Phys.\ Rev.\ D {\bf 70}, 077701 (2004)
%  doi:10.1103/PhysRevD.70.077701
  [hep-ph/0403004].

\bibitem{Fox:2011fx} 
  P.~J.~Fox, R.~Harnik, J.~Kopp and Y.~Tsai,
  %``LEP Shines Light on Dark Matter,''
  Phys.\ Rev.\ D {\bf 84}, 014028 (2011)
%  doi:10.1103/PhysRevD.84.014028
  [arXiv:1103.0240 [hep-ph]].
  %%CITATION = doi:10.1103/PhysRevD.84.014028;%%

\bibitem{Chae:2012bq} 
  Y.~J.~Chae and M.~Perelstein,
  %``Dark Matter Search at a Linear Collider: Effective Operator Approach,''
  JHEP {\bf 1305}, 138 (2013)
%  doi:10.1007/JHEP05(2013)138
  [arXiv:1211.4008 [hep-ph]].
  %%CITATION = doi:10.1007/JHEP05(2013)138;%%

\bibitem{Essig:2012yx} 
  R.~Essig, A.~Manalaysay, J.~Mardon, P.~Sorensen and T.~Volansky,
  %``First Direct Detection Limits on sub-GeV Dark Matter from XENON10,''
  Phys.\ Rev.\ Lett.\  {\bf 109}, 021301 (2012)
%  doi:10.1103/PhysRevLett.109.021301
  [arXiv:1206.2644 [astro-ph.CO]].
  %%CITATION = doi:10.1103/PhysRevLett.109.021301;%%

\bibitem{Essig:2015cda} 
  R.~Essig, M.~Fernandez-Serra, J.~Mardon, A.~Soto, T.~Volansky and T.~T.~Yu,
  %``Direct Detection of sub-GeV Dark Matter with Semiconductor Targets,''
  arXiv:1509.01598 [hep-ph].
  %%CITATION = ARXIV:1509.01598;%%
  
  %\cite{Hochberg:2015pha}
\bibitem{Hochberg:2015pha} 
  Y.~Hochberg, Y.~Zhao and K.~M.~Zurek,
  %``Superconducting Detectors for Super Light Dark Matter,''
  arXiv:1504.07237 [hep-ph].
  %%CITATION = ARXIV:1504.07237;%%
  %8 citations counted in INSPIRE as of 12 Dec 2015
  
  %\cite{Clowe:2003tk}
\bibitem{Clowe:2003tk}
  D.~Clowe, A.~Gonzalez and M.~Markevitch,
  %``Weak lensing mass reconstruction of the interacting cluster 1E0657-558: Direct evidence for the existence of dark matter,''
  Astrophys.\ J.\  {\bf 604}, 596 (2004)
  [astro-ph/0312273].
  %%CITATION = ASTRO-PH/0312273;%%
  %153 citations counted in INSPIRE as of 14 Feb 2014

%\cite{Markevitch:2003at}
\bibitem{Markevitch:2003at}
  M.~Markevitch, A.~H.~Gonzalez, D.~Clowe, A.~Vikhlinin, L.~David, W.~Forman, C.~Jones and S.~Murray {\it et al.},
  %``Direct constraints on the dark matter self-interaction cross-section from the merging galaxy cluster 1E0657-56,''
  Astrophys.\ J.\  {\bf 606}, 819 (2004)
  [astro-ph/0309303].
  %%CITATION = ASTRO-PH/0309303;%%
  %164 citations counted in INSPIRE as of 14 Feb 2014

 %\cite{Randall:2007ph}
\bibitem{Randall:2007ph}
  S.~W.~Randall, M.~Markevitch, D.~Clowe, A.~H.~Gonzalez and M.~Bradac,
  %``Constraints on the Self-Interaction Cross-Section of Dark Matter from Numerical Simulations of the Merging Galaxy Cluster 1E 0657-56,''
  Astrophys.\ J.\  {\bf 679}, 1173 (2008)
  [arXiv:0704.0261 [astro-ph]].
  %%CITATION = ARXIV:0704.0261;%%
  %127 citations counted in INSPIRE as of 10 Feb 2014

%\cite{Rocha:2012jg,Zavala:2012us,Peter:2012jh}
\bibitem{Rocha:2012jg}
  M.~Rocha, A.~H.~G.~Peter, J.~S.~Bullock, M.~Kaplinghat, S.~Garrison-Kimmel, J.~Onorbe and L.~A.~Moustakas,
  %``Cosmological Simulations with Self-Interacting Dark Matter I: Constant Density Cores and Substructure,''
  Mon.\ Not.\ Roy.\ Astron.\ Soc.\  {\bf 430}, 81 (2013)
  [arXiv:1208.3025 [astro-ph.CO]].
  %%CITATION = ARXIV:1208.3025;%%
  %61 citations counted in INSPIRE as of 14 Feb 2014

  %\cite{Zavala:2012us}
\bibitem{Zavala:2012us}
  J.~Zavala, M.~Vogelsberger and M.~G.~Walker,
  %``Constraining Self-Interacting Dark Matter with the Milky Way's dwarf spheroidals,''
  Monthly Notices of the Royal Astronomical Society: Letters {\bf 431}, L20 (2013)
  [arXiv:1211.6426 [astro-ph.CO]].
  %%CITATION = ARXIV:1211.6426;%%
  %25 citations counted in INSPIRE as of 14 Feb 2014

%\cite{Peter:2012jh}
\bibitem{Peter:2012jh} 
  A.~H.~G.~Peter, M.~Rocha, J.~S.~Bullock and M.~Kaplinghat,
  %``Cosmological Simulations with Self-Interacting Dark Matter II: Halo Shapes vs. Observations,''
  Mon.\ Not.\ Roy.\ Astron.\ Soc.\  {\bf 430}, 105 (2013)
%  doi:10.1093/mnras/sts535
  [arXiv:1208.3026 [astro-ph.CO]].
  %%CITATION = doi:10.1093/mnras/sts535;%%
  %132 citations counted in INSPIRE as of 10 Dec 2015

%\cite{Vogelsberger:2012ku}
\bibitem{Vogelsberger:2012ku} 
  M.~Vogelsberger, J.~Zavala and A.~Loeb,
  %``Subhaloes in Self-Interacting Galactic Dark Matter Haloes,''
  Mon.\ Not.\ Roy.\ Astron.\ Soc.\  {\bf 423}, 3740 (2012)
%  doi:10.1111/j.1365-2966.2012.21182.x
  [arXiv:1201.5892 [astro-ph.CO]].
  %%CITATION = doi:10.1111/j.1365-2966.2012.21182.x;%%
  %137 citations counted in INSPIRE as of 10 Dec 2015
  
  %\cite{Kaplinghat:2015aga}
\bibitem{Kaplinghat:2015aga} 
  M.~Kaplinghat, S.~Tulin and H.~B.~Yu,
  %``Dark Matter Halos as Particle Colliders: A Unified Solution to Small-Scale Structure Puzzles from Dwarfs to Clusters,''
  arXiv:1508.03339 [astro-ph.CO].
  %%CITATION = ARXIV:1508.03339;%%
  %6 citations counted in INSPIRE as of 10 Dec 2015
  
  %\cite{Spergel:1999mh}
\bibitem{Spergel:1999mh} 
  D.~N.~Spergel and P.~J.~Steinhardt,
  %``Observational evidence for selfinteracting cold dark matter,''
  Phys.\ Rev.\ Lett.\  {\bf 84}, 3760 (2000)
%  doi:10.1103/PhysRevLett.84.3760
  [astro-ph/9909386].
  %%CITATION = doi:10.1103/PhysRevLett.84.3760;%%
  %771 citations counted in INSPIRE as of 10 Dec 2015
  
  %\cite{deBlok:2009sp}
\bibitem{deBlok:2009sp} 
  W.~J.~G.~de Blok,
  %``The Core-Cusp Problem,''
  Adv.\ Astron.\  {\bf 2010}, 789293 (2010)
%  doi:10.1155/2010/789293
  [arXiv:0910.3538 [astro-ph.CO]].
  %%CITATION = doi:10.1155/2010/789293;%%
  %122 citations counted in INSPIRE as of 10 Dec 2015
  
  %\cite{BoylanKolchin:2011de}
\bibitem{BoylanKolchin:2011de} 
  M.~Boylan-Kolchin, J.~S.~Bullock and M.~Kaplinghat,
  %``Too big to fail? The puzzling darkness of massive Milky Way subhaloes,''
  Mon.\ Not.\ Roy.\ Astron.\ Soc.\  {\bf 415}, L40 (2011)
  [arXiv:1103.0007 [astro-ph.CO]].
  %%CITATION = ARXIV:1103.0007;%%
  %296 citations counted in INSPIRE as of 10 Dec 2015

%\cite{Kaplinghat:2015aga}
\bibitem{Kaplinghat:2015aga} 
  M.~Kaplinghat, S.~Tulin and H.~B.~Yu,
  %``Dark Matter Halos as Particle Colliders: A Unified Solution to Small-Scale Structure Puzzles from Dwarfs to Clusters,''
  arXiv:1508.03339 [astro-ph.CO].
  %%CITATION = ARXIV:1508.03339;%%
  %6 citations counted in INSPIRE as of 10 Dec 2015

\end{thebibliography}
\end{document}